\documentstyle[aps,prl,twocolumn,floats,amssymb,psfig,epsf]{revtex}

\begin{document}
\draft
\twocolumn[\hsize\textwidth\columnwidth\hsize\csname@twocolumnfalse\endcsname
\title{
Chiral exponents in frustrated spin models with noncollinear order
}
\author{Andrea Pelissetto,$\,^a$\cite{AP-email} Paolo Rossi,$\,^b$\cite{PR-email} 
and Ettore Vicari$\,^b$\cite{EV-email} }
\address{$^a$
Dipartimento di Fisica dell'Universit\`a di Roma I
and I.N.F.N., I-00185 Roma, Italy 
}
\address{$^b$ 
Dipartimento di Fisica dell'Universit\`a and I.N.F.N., 
Via Buonarroti 2, I-56127 Pisa, Italy.
}

\date{\today}
\maketitle

\begin{abstract}
We compute the chiral critical exponents for the chiral transition in
frustrated two- and three-component spin systems with noncollinear
order, such as stacked triangular antiferromagnets (STA). 
For this purpose, we calculate and analyze the six-loop field-theoretical expansion of the
renormalization-group function associated with the chiral operator.
The results are in satisfactory agreement with those obtained in the 
recent experiment on 
the $XY$ STA CsMnBr$_3$ reported  by V. P. Plakhty et al., Phys. Rev. Lett. 85, 3942 (2000),
providing further support for the continuous nature 
of the chiral transition.
\end{abstract}


]


The critical behavior of frustrated spin systems with noncollinear order
is still a controversial issue,
field-theoretical (FT) methods, Monte Carlo (MC) simulations,
and experiments providing contradictory results in many cases.
At present there is no agreement on the nature of  the
phase transition, and in particular on the existence of a new chiral
universality class \cite{Kawamura-88}.
See, e.g., the recent works \cite{PKVMW-00,TDM-00,PRV-01} and 
Refs. \cite{CP-97,Kawamura-98,PV-review} for reviews.  

In magnets noncollinear order is due to frustration that may arise
either because of the special geometry of the lattice, or from the competition 
of different kinds of interactions.
Typical examples of systems of the first type are 
two- and three-component antiferromagnets on 
stacked triangular lattices \cite{STA}. 
Their behavior at the chiral transition may  be modeled by a
short-ranged Hamiltonian for $N$-component spin variables $S_a$,
defined on a stacked triangular lattice as
\begin{equation}
{\cal H}_{\rm STA} = 
     - J\,\sum_{\langle ij\rangle_{xy}}  \vec{S}(i) \cdot \vec{S}(j) -
       J'\,\sum_{\langle ij\rangle_z}  \vec{S}(i) \cdot \vec{S}(j),
\label{latticeSTA}
\end{equation}
where $J<0$, the first sum is over nearest-neighbor pairs within
triangular layers, and the second one is over 
orthogonal interlayer nearest neighbors.
Frustration due to the competition of interactions is realized in
helimagnets.

\begin{figure}
\vspace*{0truecm} \hspace*{-0.2cm}
\centerline{\psfig{width=6truecm,angle=0,file=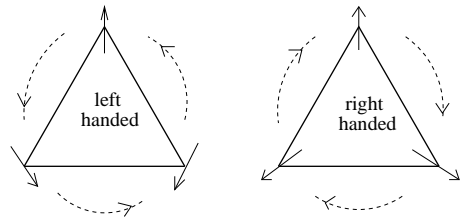}}
\vspace*{0.3cm}
\caption{
The ground-state configuration of three $XY$ spins on a triangle
coupled antiferromagnetically. 
}
\label{chiralityfig}
\end{figure}

In these models frustration is partially released by mutual spin canting
and the degeneracy of the ground state is limited to global O($N$) 
spin rotations and reflections. 
At criticality one expects 
a breakdown of the symmetry from O($N$) in the high-temperature phase
to O($N-2$) in the low-temperature phase,  implying a matrix-like order parameter.
In particular, the ground state of the $XY$ systems  
shows the 120$^o$ structure of Fig.~\ref{chiralityfig},
and it is $Z_2$ chirally degenerate  
according to whether the noncollinear spin configuration is right- or
left-handed.
The chiral degrees of freedom are related to 
the local quantity~\cite{Kawamura-88}
\begin{equation}
C_{ab} \propto \sum_{\vartriangle} S_a(i)S_b(j) - S_b(i) S_a(j)
\label{chirality}
\end{equation}
where the summation runs over the three bonds of the given triangle. 
The definition of $C_{ab}$ can be straightforwardly generalized to the case
of $N$-component spins.

Many experiments are consistent with a second-order phase transition 
belonging to a new (chiral)  universality class
(see,  e.g., Refs. \cite{CP-97,Kawamura-98} for reviews).
Further experimental evidence in favor of a chiral continuous transition
has been recently reported in Ref.~\cite{PKVMW-00},
showing  the simultaneous occurrence
of spin and chiral order in the $XY$ stacked triangular antiferromagnet
(STA)  CsMnBr$_3$.

On the theoretical side the issue has been controversial.
MC simulations \cite{Kawamura-92,BBLJ-94,PM-94,MPC-94,LD-94,BLD-96,LS-98}
(see Refs.~\cite{Kawamura-98,PV-review} for reviews)
have not been conclusive in setting the question. Simulations of  
STA's are consistent with continuous transitions, but 
with critical exponents that are not in a satisfactory quantitative agreement.
In Ref.~\cite{PM-94} the results for the $XY$ STA are interpreted
as an evidence for  mean-field tricritical behavior.
Moreover, MC investigations~\cite{LS-98} of special lattice spin
systems, that on the basis of their symmetry should belong to the 
chiral universality class, show  clearly a first-order transition.

In a recent Letter~\cite{TDM-00}  the issue has been studied by
a continuous renormalization-group (RG) approach (see also Refs.~\cite{TDM-01,Zumbach-94}).
The results favor a first-order transition, since no evidence of 
stable fixed points is found.
According to this first-order transition picture,
the apparent continuous critical phenomena observed in experiments
are interpreted as first-order transitions,
weak enough to effectively appear
as second-order ones.
Note however that the practical implementation of this 
method requires an approximation and/or truncations of the effective action.
So these studies may not be conclusive.

\begin{figure}
\vspace*{0truecm} \hspace*{-0.2cm}
\centerline{\psfig{width=6truecm,angle=0,file=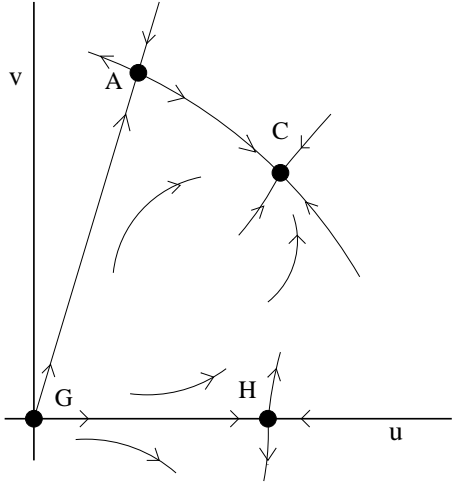}}
\vspace*{0.3cm}
\caption{RG flow in the quartic couplings $(u,v)$ plane for $N=2,3$.
It shows the stable chiral fixed point denoted by $C$,
and the unstable antichiral ($A$), O(2$N$) Heisenberg ($H$) and
Gaussian ($G$) ones.
}
\label{rgflow}
\end{figure}

FT studies of systems with noncollinear order are based on the
Landau-Ginzburg-Wilson 
O($N$)$\times$O(2)-symmetric  Hamiltonian~\cite{Kawamura-88,Kawamura-98}
\begin{eqnarray}
{\cal H} = \int d^3 x 
&& \Bigl\{ {1\over2}
      \sum_{a} \left[ (\partial_\mu \phi_{a})^2 + r \phi_{a}^2 \right] 
+ {1\over 4!}u_0 ( \sum_a \phi_a^2 )^2  \nonumber\\
&&  + {1\over 4!}  v_0 
\sum_{a,b} \left[ ( \phi_a \cdot \phi_b)^2 - \phi_a^2\phi_b^2\right]
             \Bigr\},\label{LGWH}
\end{eqnarray}
where $\phi_a$ ($1\leq a\leq 2$) are two sets of $N$-component
vectors. 
Frustrated $XY$ and Heisenberg spin systems with noncollinear
ordering, such as STA's,
are described respectively by the $N=2$ and $N=3$ case with $v_0>0$.
The presence of a stable chiral fixed point, conjectured 
by Kawamura \cite{Kawamura-88,Kawamura-98},
has been recently confirmed by 
the analysis of the 
perturbative six-loop series 
in the framework of the fixed-dimension expansion 
\cite{PRV-01,earlierstudies}.
As sketched in Fig.~\ref{rgflow},  a stable chiral fixed point $C$
appears for both $XY$ and Heisenberg cases. 
The critical exponents 
characterizing the stable chiral fixed point
turn out to be in satisfactory agreement with experiments.
Note that in this RG picture
first-order transitions are still possible  for systems that 
are outside the attraction domain of the chiral fixed point.
In this case, the RG flow runs away to a 
first-order transition. This may explain some experiments
(for example those for the CsCuCl$_3$ compound, see, e.g.,
Refs.~\cite{CP-97,Kawamura-98}) 
and MC studies for special lattice systems \cite{LS-98},  
where first-order transitions are observed. 

Beside the conventional critical exponents $\beta$, $\gamma$, $\nu$,
etc... related to the standard spin order, 
one may consider additional critical
exponents related to  the behavior of the chiral degrees of
freedom.  
If spin and chiral order occur simultaneously, one expects $\nu_c=\nu$ where
$\nu_c$ is the exponent associated with 
the correlation length defined from the
chiral correlation function.
Introducing a chiral external field $h_c$ coupled with the chirality $C_{ab}$, 
one may write the singular part of the free energy as \cite{Kawamura-88}
\begin{equation}
F_{\rm sing} \propto t^{2-\alpha} f\left( h/t^\Delta,h_c/t^{\phi_c} \right),
\label{freeen}
\end{equation}
where $t$ is the reduced temperature,
$\Delta = \beta + \gamma$, and $\phi_c$ is the chiral crossover
exponent. Then, differentiating with respect to $h_c$, one may obtain
the RG relations 
\begin{equation}
\beta_c = 3\nu - \phi_c,\qquad \gamma_c = 2 \phi_c - 3\nu,
\label{scalrel}
\end{equation}
where $\beta_c$ and  $\gamma_c$ describe respectively the critical behavior of 
the average  chirality and of the chiral susceptibility.

The first esperimental estimates of the chiral exponents
$\phi_c$ and $\beta_c$ have been recently reported in 
Ref.~\cite{PKVMW-00}
for the transition of the $XY$ STA CsMnBr$_3$:
\begin{equation}
\phi_c=1.28(7), \qquad
\beta_c=0.44(2), \label{expres} 
\end{equation}
measured respectively in the high- and low-temperature phase.
On the theoretical side, there are a few MC results for the STA spin
models (\ref{latticeSTA}), and very little from field-theoretical approaches. 
The chiral exponents  have  been only
computed  to O($1/N$) and $O(\epsilon)$ in the corresponding
expansion frameworks \cite{Kawamura-88}. However, these results do not allow 
a quantitative comparison, essentially for two reasons:
because the series are too short and, most importantly,
as discussed in Ref.~\cite{PRV-01-2}, the chiral fixed point for the $XY$
and Heisenberg cases 
is not  analytically connected with the one
found in the large-$N$ and small-$\epsilon$ region.
In order to obtain results that can be compared with experiments,
one should compute them directly for $d=3$ and for the 
number of components of interest, i.e. $N=2,3$.

In this Letter we compute the chiral exponents using the
FT approach of Ref.~\cite{PRV-01},
i.e. by computing and analyzing the six-loop perturbative expansion
of the chiral RG functions.
In the fixed-dimension FT approach one 
performs an expansion in powers of appropriately defined 
zero-momentum quartic couplings (see, e.g., Ref.~\cite{PV-review} and
references therein). 
In order to obtain estimates of the universal critical quantities, 
the perturbative series are resummed
and then evaluated at the fixed-point values of the couplings. 
The comparison with the experimental results (\ref{expres}) will 
represent a highly nontrivial check of the FT description of the
transition and of the Kawamura's conjecture that 
these systems undergo continuous transitions belonging
to distinct chiral universality classes.

In order to compute the universal quantities characterizing
the critical behavior in the high-temperature phase, 
one introduces a set of zero-momentum conditions 
for the (one-particle irreducible) two-point and four-point
correlation functions (see, e.g., Ref.~\cite{PV-review} for details),
which relate 
the zero-momentum quartic couplings $u$ and $v$ and the mass scale $m$
to the corresponding Hamiltonian parameters
$u_0$, $v_0$ and $r$.
In particular,
\begin{equation}
\Gamma^{(2)}_{ai,bj}(p) = \delta_{ab}\delta_{ij} Z_\phi^{-1} \left[ m^2+p^2+O(p^4)\right].
\end{equation}
In addition, one defines the function  $Z_t$ through the relation
$\Gamma^{(1,2)}_{ai,bj}(0) = \delta_{ab}\delta_{ij} Z_t^{-1}$,
where $\Gamma^{(1,2)}$ is the (one-particle irreducible)
two-point function with an insertion of $\case{1}{2}\phi^2$.
The fixed points of the theory are given by 
the common  zeros $u^*$, $v^*$ of the $\beta$-functions
\begin{equation}
\beta_u(u,v) = m{\partial u\over \partial m},
\qquad
\beta_v(u,v) = m{\partial v\over \partial m},
\end{equation}
calculated keeping $u_0$ and $v_0$ fixed.
The critical behavior is determined by the stable fixed point
of the theory. The analysis of the six-loop expansion of the
$\beta$-functions provided a rather robust evidence of the existence 
of a stable fixed point \cite{PRV-01}  as shown in Fig.~\ref{rgflow}.
The critical exponents $\eta$ and $\nu$ are then derived 
by evaluating the RG functions
\begin{equation}
\eta_\phi(u,v) = {\partial \ln Z_\phi\over \partial \ln m},\qquad
\eta_t(u,v) = {\partial \ln Z_t\over  \partial \ln m}
\end{equation}
at the chiral fixed point $u^*$, $v^*$.
The resulting exponents are 
$\nu=0.57(3)$,  $\eta = 0.09(1)$, $\gamma=1.13(5)$  for the $XY$ case, and
$\nu=0.55(3)$, $\eta=0.10(1)$, $\gamma=1.06(5)$  for the Heisenberg
case \cite{expXYH},
which are in substantial  agreement with the experimental results.

In order to evaluate the chiral exponents, we consider the operator 
\begin{equation}
C_{ckdl}(x) = \phi_{ck}(x) \phi_{dl}(x) - \phi_{cl}(x) \phi_{dk}(x),
\label{chiralop}
\end{equation} 
and define a related
renormalization function $Z_c$ from the one-particle irreducible
two-point function $\Gamma^{(c,2)}$ with an insertion of the operator
$C_{ai,bj}$, i.e.
\begin{equation}
\Gamma^{(c,2)}(0)_{ai,bj,ckdl} = Z_c^{-1} \; T_{abcd,ijkl}
\end{equation}
where
\begin{equation}
T_{abcd,ijkl} = 
\left( \delta_{ac}\delta_{bd} -\delta_{ad}\delta_{bc} \right)
\left( \delta_{ik}\delta_{jl} - \delta_{il}\delta_{jk} \right)
\end{equation}
so that $Z_c(0,0)=1$.
Then, we compute  the RG function 
\begin{equation}
\eta_c(u,v) = {\partial \ln Z_c \over \partial \ln m}
= \beta_u {\partial \ln Z_c \over \partial u} +
\beta_v {\partial \ln Z_c \over \partial v} ,
\end{equation}
and its value  $\eta_c$ at $u=u^*$, $v=v^*$, where 
$u^*$, $v^*$ is the position of  the stable chiral fixed point \cite{PRV-01}.
Finally, the RG scaling relation
\begin{equation}
\phi_c = \left( 2 + \eta_c - \eta\right)\nu
\end{equation}
allows us to determine $\phi_c$.

We computed $\Gamma^{(c,2)}(0)$ to six loops. 
The calculation is rather cumbersome, since it requires
the evaluation of 563 Feynman  diagrams.
We handled it with a symbolic manipulation program, which  generates the diagrams 
and computes the symmetry and group factors of each of them.
We used the numerical results compiled in Ref.~\cite{NMB-77}
for the integrals associated with each diagram.
The resummation of the series was performed using the method outlined
in Refs.~\cite{CPV-00,PRV-01}.
The very lengthy expression of the 
six-loop expansion of $\eta_c(u,v)$, 
details of its calculation, and its analysis will be reported elsewhere.
The results of our analysis are
\begin{eqnarray}
&\phi_c =  1.43(4) \qquad &{\rm for} \quad XY ,\\
&\phi_c =  1.27(4) \qquad &{\rm for} \quad {\rm Heisenberg} .
\end{eqnarray}
The errors are indicative of the spread of the results yielded by the
analysis when changing the resummation parameters
and varying the location of the chiral fixed
point within the range reported in Ref.~\cite{PRV-01}.
Using the RG relations (\ref{scalrel})
and the estimates of $\nu$ \cite{PRV-01}, one may also derive corresponding results
for the other chiral exponents, obtaining for example
\begin{eqnarray}
&\beta_c =  0.28(10) \qquad &{\rm for} \quad XY ,\\
&\beta_c =  0.38(10) \qquad &{\rm for} \quad {\rm Heisenberg}.
\end{eqnarray}
We may compare these results with the 
experimental ones (\ref{expres}). Our estimate of 
$\phi_c$ is somewhat higher than the estimate  (\ref{expres})
while the estimate of $\beta_c$ is correspondingly
somewhat lower. In any case, the difference is of the 
order of one combined error bar. We may also compare 
our results to the available MC estimates for the $XY$ STA spin model,
that are
$\beta_c=0.45(2)$, $\gamma_c=0.77(5)$, $\phi_c=1.22(6)$ 
from Ref.~\cite{Kawamura-92},
and $\beta_c=0.38(2)$, $\gamma_c=0.90(9)$, $\phi_c=1.28(10)$
from Ref.~\cite{PM-94} 
(in this work a mean-field tricritical behavior is conjectured for the transition).
These results are close
to the experimental ones and thus show the same  
deviations with respect to our FT results.

For $N=3$ we can compare our results with 
the MC ones for the three-component STA spin model, that are
$\beta_c=0.55(4)$, $\gamma_c=0.72(8)$, $\phi_c=1.27(9)$
from Ref.~\cite{Kawamura-92}
and $\beta_c=0.50(2)$ $\gamma_c=0.82(4)$ and $\phi_c=1.32(5)$ 
from Ref.~\cite{MPC-94}. 
The FT estimate of $\phi_c$ is in perfect agreement with 
the MC results, while the estimate of $\beta_c$ is 
somewhat lower, although compatible within error
bars. Apparently, this is due to the fact that our 
estimate of $\nu$ is somewhat lower than 
those obtained in MC simulations.

In conclusion, the FT results are in satisfactory agreement 
with the experimental and MC estimates. This is a 
nontrivial check of the FT approach, shows its 
predictive power in spite of the fact that the 
perturbative series are not Borel summable---still 
we take into account the leading diverging 
behavior, see Ref. \cite{PRV-01}---and strengthens
the evidence for the continuous nature of 
the chiral transition in XY and Heisenberg STA's.


\begin{references}
\bibitem{STA}
Examples of STA are 
${\rm ABX}_3$-type compounds, where A denotes elements such as Cs and Rb, B stands for magnetic
ions such as Mn, Cu, Ni, and Co, and X for halogens as Cl, Br, and I.
\bibitem{earlierstudies}
Earlier studies based on three-loop series in the fixed-dimension \cite{AS-94}
and $\epsilon$-expansion \cite{ASV-95} frameworks did not find any
evidence for new stable chiral fixed points.
\bibitem{expXYH}
The critical exponents found for the chiral $XY$ and Heisenberg
universality classes are considerably different from 
those of the standard $XY$ and Heisenberg ones. 
We mention their recent estimates \cite{CHPRV-01}: 
$\nu=0.67155(27)$, $\eta=0.0380(4)$, $\gamma=1.3177(5)$ for the $XY$ universality class,
and  $\nu=0.7112(5)$, $\eta=0.0375(5)$, $\gamma=1.3960(9)$ for the Heisenberg one.
\end{references}
\end{document}